# ION-AEROSOL INTERACTIONS IN THE LOWER ATMOSPHERE OF VENUS

K.L. Aplin

**Space Science and Technology Department, Rutherford Appleton Laboratory, Chilton, Didcot, Oxon OX11 0QX UK**

### Introduction: the atmospheric electrical environment on Venus

All planetary atmospheres are electrified to some extent by cosmic ray ionization, and Venus is no exception. There is increasing awareness that ion-aerosol interactions could modulate terrestrial radiative processes, and this possibility will be investigated for the Venusian atmosphere. The likelihood of a Venusian global atmospheric electric circuit (defined below) will also be discussed.

Venus' atmosphere is dense enough for ions and electrons formed by cosmic rays to rapidly produce ion clusters[1]. Common predicted positive species are $H_3O^+(H_2O)_n$ (n = 3 or 4), $H_3O^+(SO_2)$ and $H_3O^+(H_2O)(SO_2)$. Sulphate species dominate predicted negative ion species, and free electrons are also expected above 60km. Because of the ubiquity of the cloud cover (at ~50-70km), attachment to cloud particles is an important global ion loss mechanism, significantly reducing charged particle concentrations[1].

### Could there be a global electric circuit on Venus?

A global atmospheric at electric circuit needs, at the minimum, electrostatic discharge or charged particle precipitation balanced by current carried by mobile charged particles between a conductive ionosphere and surface[2]. A global circuit causes transport of charged particles between different parts of the atmosphere. At the moment, the existence of a Venusian global electric circuit seems unlikely due to the lack of proof for electrostatic discharges[3]. If lightning is unambiguously observed, then there is a good basis for the existence of a global circuit because of the presence of ions and electrons and a conducting ionosphere.

### Electrification and particle formation

A potentially important mechanism for cloud formation at the bottom of the cloud layer is vapour condensing onto aerosol particles[4]. The atmosphere is supersaturated with respect to sulphuric acid from 40km upwards, where $H_2SO_4$ vapour can condense onto hydrated sulphuric acid particles. As Venusian atmospheric ions are predicted to be sulphuric acid hydrates, appropriate conditions may exist for aerosols to form by ion-induced nucleation. The supersaturation required for ions to grow into ultrafine droplets by direct condensation can be determined using the Thompson equation[5]. This equation describes the equilibrium saturation ratio needed for ion-induced nucleation to become energetically favourable. The equilibrium condition is defined at a saturation ratio $S$, where $r$ is radius, $\rho$ fluid density, $M$ the mass of the molecule, $q$ charge, $\gamma_T$ the surface tension, $k_B$ Boltzmann's constant, $T$ temperature, $r_o$ the initial radius (all in SI units), and $\varepsilon_r$ relative permittivity:

$$\ln S = \frac{M}{k_B T \rho}\left[\frac{2\gamma_T}{r} - \frac{q^2}{32\pi^2\varepsilon_0 r^4}\left(1-\frac{1}{\varepsilon_r}\right)\right]. \qquad (1)$$

(1) can be used to assess if condensation of gaseous $H_2SO_4$ onto ions is likely to occur in the lower cloud-forming regions at ~40km. To calculate $S$, $H_2SO_4$ concentrations and supersaturations determined from microwave absorption measurements are used[6]. Temperatures were obtained from a model atmosphere[1]. Surface tension, density and dielectric constant for ~100% sulphuric acid at 400K were estimated and extrapolated from existing data[2]. (1) was then used to compute the saturation ratio needed for nucleation onto ions with 1, 2 and 5 electronic charges at a temperature $T$, for the latitudes at which the saturation ratios were measured.

At 88°S, supersaturation (SS) is expected at ~47km and 43km, but at the lower altitude the higher temperatures reduce the SS needed for activation. Figure 1 shows the solution of (1) at 395K. SS is never high enough to activate particles with one electronic charge, but doubly charged particles of radius 1nm can nucleate at SS=7%, and particles with 5 charges of radius 2nm are activated at SS=1-2%, relatively easily attained SS levels[6]. The number of doubly charged particles can be estimated from the aerosol charge distribution arising from ion-aerosol attachment processes. The steady state charge distribution of a monodisperse aerosol population, represented as the ratio of the number of particles with charge $j$, $N_j$, to the number of neutral particles, $N_0$, can be given by

$$\cdot \frac{N_j}{N_0} = x^j \frac{8\pi\varepsilon_0 akT}{je^2} \sinh\left[\frac{je^2}{8\pi\varepsilon_0 akT}\right] \exp\left[\frac{-j^2 e^2}{8\pi\varepsilon_0 akT}\right] \qquad (2)$$

where $j$ is the electronic charge, $a$ the mean aerosol radius, and $x$ is the "ion asymmetry factor"[7]. The charge distribution can be calculated for T=395K and typical ion properties at 40km. The mobility of positive ions was assumed to be 7% higher than negative ions, with no free electrons below 60km[1]. Positive and negative ions were assumed to exist in equal concentrations. The aerosol diameter at 40km was assumed to be 0.25μm[1]. Substituting for $a$, $T$ and $x$ in (2) indicates that 84% of aerosols in this region carry some charge, with the most common single charge state being +1. The mean charge is +0.2, but, across the estimated charge distribution, Figure 2, $j \geq 2$ is relatively common with 54% of Venusian aerosols carry enough charge for ion-induced nucleation at supersaturations of 7%, and 7% of the aerosol population carries enough charge for nucleation at 1-2% supersaturation. These estimates suggest that it may be possible for gaseous sulphuric acid to condense onto ions and form aerosol particles.

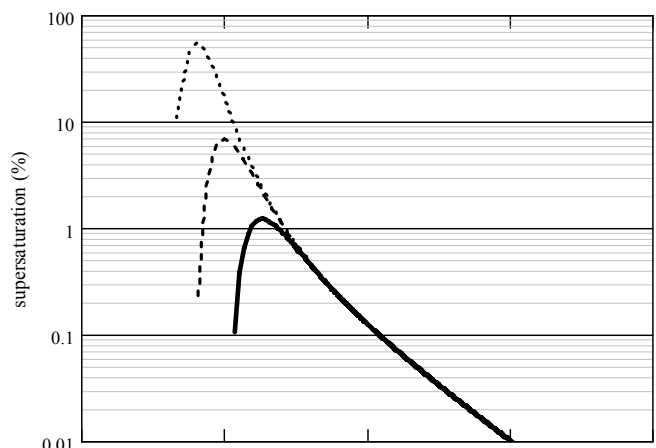

Figure 1 Supersaturations required for $H_2SO_4$ nucleation onto ions with 1, 2 and 5 electronic charges for conditions at 88ºS, 43 km.

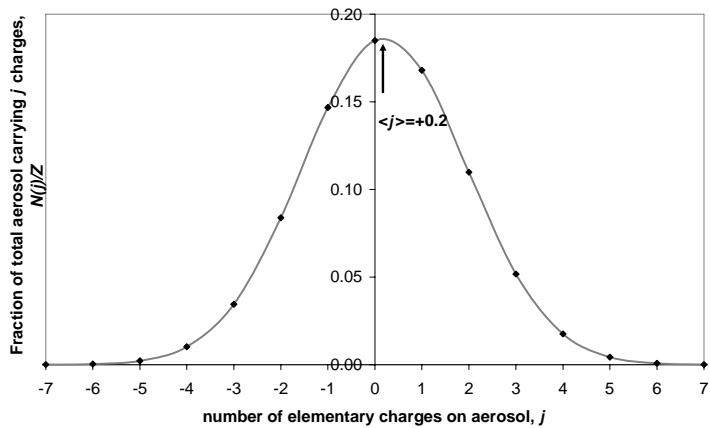

Figure 2 Steady state charge distribution on Venus aerosols with mean diameter of 0.25 μm at 395 K.

**Conclusions**
- Venus almost certainly has active atmospheric electrical physics, though further measurements are needed to determine whether a global circuit exists. Because of cluster-ion formation in a dense gas, a global circuit could affect the location of charged particles and thus influence cloud or aerosol particle formation and lifetime. From a comparative planetology viewpoint, a Venusian global circuit with its permanent cloud would be interesting to compare to the effect of variable cloud cover on Earth's global circuit.
- Formation of aerosol particles by heterogeneous nucleation onto ions may be possible. If this does occur then cosmic rays modulated by the solar cycle could potentially affect climate.
- Any charged particle and aerosol measurements should be simultaneous so that ion-aerosol interactions can also be investigated. Larger charged particles should be measured, as well as the ions that contribute to air conductivity. This could be achieved with a relaxation probe, like the Huygens PWA probe, but using a higher voltage to capture larger particles.